# Coupling spin defects in hexagonal boron nitride to monolithic bullseye cavities

Johannes E. Fröch[1,*], Lesley Spencer[1,2,*], Mehran Kianinia[1,2], Daniel Totonjian[1], Minh Nguyen[1], Vladimir Dyakonov[3], Milos Toth[1,2], Sejeong Kim[4], Igor Aharonovich[1,2]

[1]School of Mathematical and Physical Sciences, University of Technology Sydney, Ultimo, New South Wales 2007, Australia

[2]ARC Centre of Excellence for Transformative Meta-Optical Systems (TMOS), University of Technology Sydney, Ultimo, New South Wales 2007, Australia

[3]Experimental Physics 6 and Würzburg-Dresden Cluster of Excellence ct.qmat, Julius Maximilian University of Würzburg, 97074 Würzburg, Germany.

[4]Department of Electrical and Electronic Engineering, University of Melbourne, Victoria, 3010, Australia

[*]These Authors contributed equally

**Corresponding**
johannes.froech@uts.edu.au; igor.aharonovich@uts.edu.au

**Abstract**
*Color centers in hexagonal boron nitride (hBN) are becoming an increasingly important building block for quantum photonic applications. Herein, we demonstrate the efficient coupling of recently discovered spin defects in hBN to purposely designed bullseye cavities. We show that the all-monolithic hBN cavity system exhibits an order of magnitude enhancement in the emission of the coupled boron vacancy spin defects. In addition, by comparative finite-difference time-domain modelling, we shed light on the emission dipole orientation, which has not been experimentally demonstrated at this point. Beyond that, the coupled spin system exhibits an enhanced contrast in optically detected magnetic resonance readout and improved signal to noise ratio. Thus, our experimental results supported by simulations, constitute a first step towards integration of hBN spin defects with photonic resonators for a scalable spin-photon interface.*

Hexagonal boron nitride, a naturally occurring van der Waals crystal, is becoming a prevalent platform to study nanophotonics and light matter interaction at the nanoscale.[1,2] Of particular importance is its ability to host optically active defects that have been recently studied as promising solid state quantum emitters.[3-16] Furthermore, several of these defects exhibit a spin - photon interface, with a clear optically detected magnetic resonance (ODMR) even at room temperature.[17-20] This class of defects that exhibit ODMR, is highly sought after for quantum sensing, quantum information and integrated quantum photonics applications.[21,22]

One specific example is the negatively charged boron vacancy ($V_B^-$) defect which has a triplet ground state, with zero field splitting of ~ 3.5 GHz and a broad emission around 810 nm.[17] While its level structure is the focus of growing research, many fundamental aspects are still unknown. These include the position of its zero phonon line and its detailed photophysical properties, as well as the fundamental reasons for the relatively low quantum efficiency (QE). [23-26] To circumvent the latter, emission enhancement must be realized, for example by coupling the $V_B^-$ defects to plasmonic or dielectric resonators. Integration of spin defects with photonic waveguides or cavities is also ultimately required to improve photon collection efficiencies and is critical for many applications in quantum photonics and quantum sensing.[27-29]

In this work we demonstrate the integration of the $V_B^-$ into a dielectric cavity - specifically a bullseye cavity.[30-35] The rationale behind the choice for this particular device type is given by three factors. First, the dipole of the $V_B^-$ emitter has not been experimentally determined. While photonic crystal cavities or plasmonic cavities are typically designed to enhance either in plane or out of plane dipole orientations, a bullseye cavity improves the collection efficiency of any dipole orientation. Second, the ensemble emission of the $V_B^-$ covers a wide spectral range from 750 nm to 850 nm, which does not narrow at cryogenic temperature. The resonance of the bullseye (several tens of nanometers) can match a broad range at the maximum of that emission, which facilitates efficient coupling to the device. Finally, the bullseye cavity is a planar device (unlike, for example, solid immersion lenses or nanoscale pillars) that dramatically enhances the collection efficiency, especially for lower numerical aperture lenses, and is therefore particularly suitable for van der Waals and 2D materials.

To enable the largest possible spatial overlap of the $V_B^-$ ensemble with the cavity fields, we employed a monolithic approach. Here, the bullseye cavity hosting the $V_B^-$ defects is fabricated entirely from hBN, as shown schematically in Figure 1(a). The monolithic hBN bullseye cavity facilitates coupling to the $V_B^-$ with a strong directionality of emission into a narrow angle of the far field, enabling improved spin readout of the emitter in hBN. The inset is a schematic of the hBN lattice hosting the $V_B^-$ spin defects.

The device fabrication is described in detail in the methods section and follows the principal steps as described in our previous studies.[36,37] Briefly, hBN was transferred from a high quality bulk crystal onto 285 nm $SiO_2$ and suitable flakes of desired thickness (~ 290 nm) were identified by optical contrast. The sample was then coated with a polymer resist and patterned via electron beam lithography. Subsequently, reactive ion etching was used to cast the pattern from the resist into the hBN. A representative optical image of a hBN flake after fabrication is shown in Figure 1(b), whereas the structures are well defined and etched entirely through to the underlying substrate, as shown by a high resolution SEM image in Figure 1(c).

For our work, we based the lattice defining parameter *a* on the second order bragg condition $a=\lambda/n_{eff}$,[38] for which we assumed an effective refractive index of $n_{eff} \sim 1.7$ at $\lambda = 800$ nm ($n_l = 2.1$)[39] with further structure defining parameters given by the central disk diameter $d_0$, and the air gap between rings *g*, with 9 rings in total. As noticeable in Figure 1(b), we tuned these parameters in order to cover a larger range of spectral resonances. In the following, the notations $BE_1$-$BE_8$ correspond to 8 different groups of bullseye cavities with different resonances that were engineered by tuning the structural parameters *a*, $d_0$, and *g*. Specifically, the scaling was set in increments of 5% smaller/larger relative to the parameters of device $BE_3$, defined by $a$=475 nm, $d_0$=950 nm, and $g$=180 nm.

After fabrication, the entire flake was homogeneously irradiated with a focused ion beam (FIB) to generate vacancies. Previous studies employed heavier ions like nitrogen or xenon to generate the $V_B^-$ defects.[40] This results in a rather shallow implantation depth. To engineer the emitters all throughout the hBN cavity (depth-wise), we used a focused hydrogen beam. (further detailed in the Methods and Supporting Information 1).

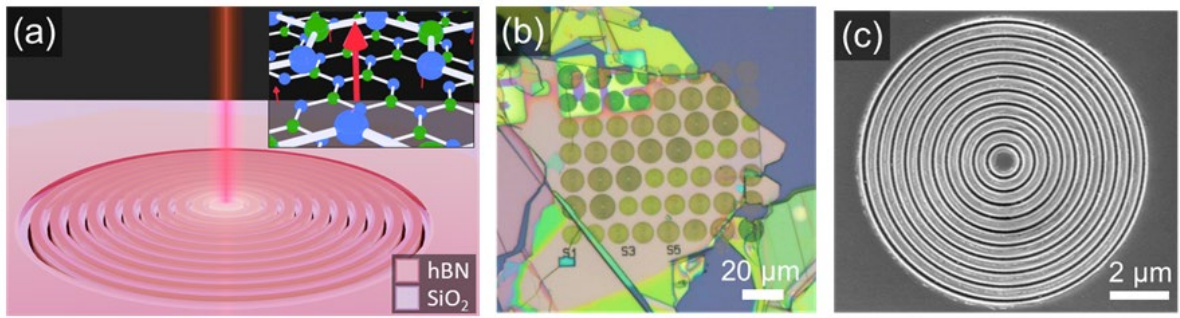

***Figure 1. The Bullseye Cavity.*** *a) A schematic of a hBN bullseye cavity on $SiO_2$ generating collimated emission into free space. The top right inset shows a schematic of the $V_B^-$ spin defect in the hBN lattice, where nitrogen and boron are depicted as blue and green spheres, respectively. The optically active $V_B^-$ spin defect is illustrated as a red arrow. b) Optical microscope image of an exfoliated hBN flake on 285 nm $SiO_2$ with an array of fabricated devices. The various colors of the flake correspond to different thicknesses. c) An SEM image of a bullseye cavity.*

We now turn to a detailed characterization of the photonic functionality. All experiments were conducted at room temperature using a lab-built confocal microscope setup with a 0.9 NA objective and a 532 nm CW laser as the excitation source. Figure 2(a) shows spectra of the bullseye cavities for each scaling (blue color scale) in comparison to the emission collected from pristine hBN (red).

Notably, we observed that devices $BE_3$ coupled to the central wavelength of the $V_B^-$ at 795 nm, while $BE_1$ and $BE_2$ show enhanced $V_B^-$ emission at shorter wavelengths of 740 nm and 765 nm, respectively. The $BE_4$ and $BE_5$ cavity modes were observed at longer wavelengths at 825 nm and 855 nm, respectively. Devices with larger scaling factor ($BE_6$ - $BE_8$) showed weakly coupled modes or no mode at all (Supporting Information 2). The quality factor, as determined by a Lorentzian fit for $BE_3$ is on the order of $\sim 100$ (Figure 2(a) inset). We note that we did not observe modes before ion beam irradiation. This is direct evidence that the light coupling to the cavity stems from $V_B^-$ ensembles as opposed to background emission (e.g. surface contamination or other luminescent defects). Furthermore, the equidistant distribution of modes with device scaling indicates that the emission is coupled to the same type of resonance (further discussed below).

We observed modes among all devices of the sets $BE_1$ - $BE_5$ with PL enhancements as outlined in Figure 2(b). On average, the bullseye cavities with a lattice constant of $a$=475nm, corresponding to $BE_3$, showed the largest gain in PL intensity (up to a factor of ~ 6.5). Even for detuned devices we still observed on average a PL enhancement by a factor of ~ 3 and for some devices up to ~ 6. Yet, in absolute numbers, devices of scaling $BE_3$ showed the largest increase in total PL intensity, due to the closest match of the mode to the center of the $V_B^-$ emission. Regardless, the fact that the enhancement occurs to be on par among different scalings shows equivalently efficient coupling of the bullseye cavity to the $V_B^-$ throughout its entire emission range. This indicates that the dipole properties of the emitter (orientation and strength) throughout the range are homogeneous.

To further characterize the system, we studied the saturation behavior of the $V_B^-$ ensembles under increasing excitation power. Measurements for bare hBN compared to the bullseye cavities are shown in Figure 2(c). Here we determined the intensity from the integrated spectrometer counts over an emission range from 775 nm to 805 nm (inset). The measured data set is fitted to the equation $I = (I_{sat}P)/(P + P_{sat})$ where $I_{sat}$ is the saturation intensity and $P_{sat}$ is a saturation power for the excitation laser. We determined saturation intensities of 242 and 123 and powers of 16.4 mW and 42.3 mW for $V_B^-$ emission coupled to a bullseye cavity and from plain hBN, respectively. The Enhancement factor for PL is ~ 6 and ~ 2 in the undersaturation and saturation regime, respectively.

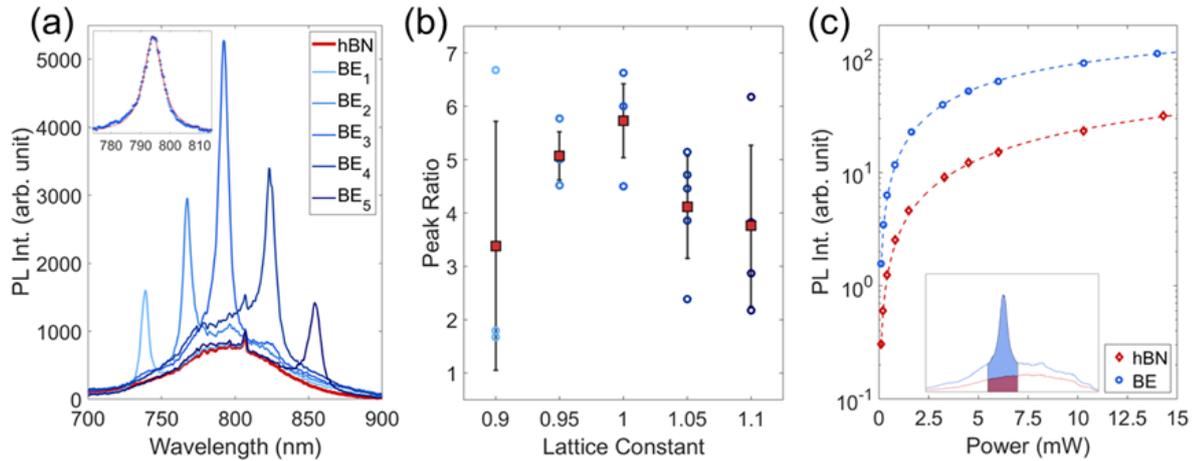

*Figure 2. PL characterization. a) Spectra of devices with varying lattice constants. The inset shows a Lorentzian fit of the mode for a device $BE_3$ with a Q-factor of ~ 100. b) Comparison of the PL enhancement for various devices relative to emission in an unstructured hBN film. Measurements of bullseye cavities are represented as blue circles. The mean and standard deviation are represented for each scaling set as red square and error bars, respectively. c) Comparison of PL saturation for $V_B^-$ emitter in a bullseye cavity (blue) and unstructured hBN (red) derived by integrating over the indicated spectral range of the $V_B^-$ (inset).*

To gain deeper insight into the enhancement mechanism, we simulated various models based on the Finite Difference Time Domain (FDTD) method (see Methods for details). Specifically, as the dipole orientation of the defect has not been experimentally determined, we consider two cases: in-plane and out-of-plane dipole relative to the hBN sheet, which couple to a TE-like and TM-like mode of the bullseye cavity, respectively. Structural parameters obtained from a SEM image of $BE_3$ are used for simulation and the results are shown in Figure 3(a). A TM (TE) mode occurs at 690 nm (785 nm), plotted as red (green) curves, while the experimental spectrum measured from the bullseye cavity shows a mode at 795 nm (blue curve). A large difference between the TM and TE resonant wavelength stems from two refractive indices (i.e. birefringence)[39] from hBN. Here, the experimental resonance wavelength nearly matches the TE mode and it is unlikely to be the TM mode even considering the fabrication imperfection that can cause the discrepancy between the simulation and experiment mode. Regardless, we emphasize further that with larger device scaling, the TE like mode shifts towards longer wavelengths (Figure 2(a)) and no further modes emerge. This is an indication of an extremely weak out-of-plane emission dipole component. Despite the identification of the emission dipole, we note that the absorption dipole may not be co-aligned,[41] which thus remains a topic of future experiments.

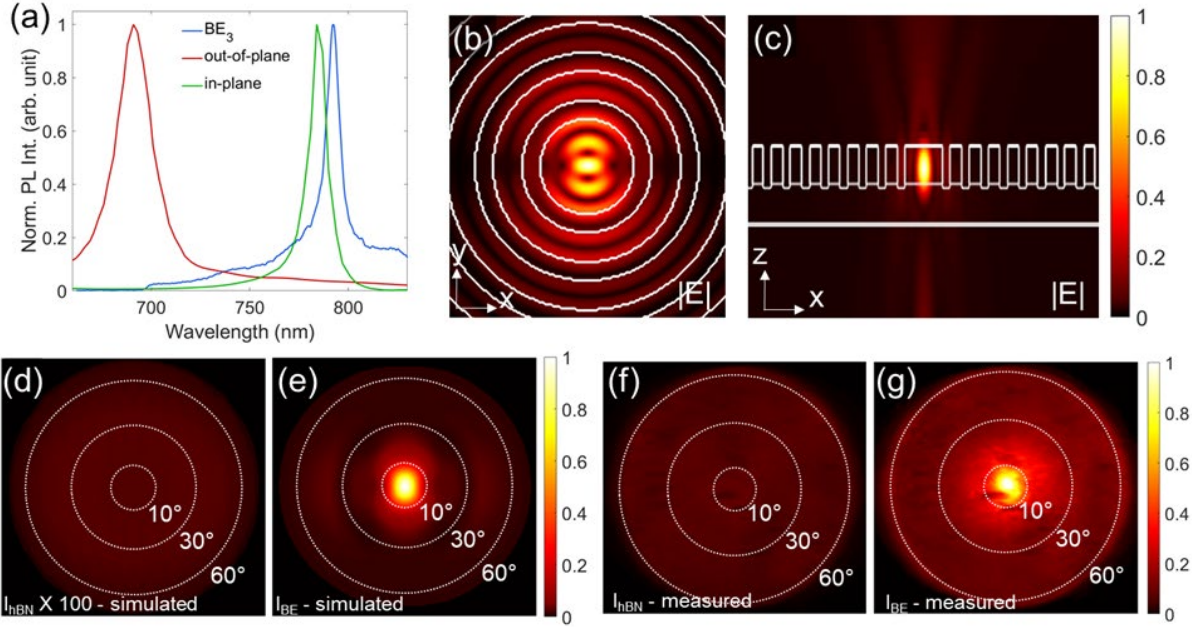

*Figure 3. Simulation and far field emission. a) Spectral comparison between device $BE_3$ (blue) to FDTD simulations assuming an out-of-plane (red) and an in-plane dipole (green). b) Top view of the field intensity inside the bullseye cavity. c) Side view of the field intensity. d) and e) show the simulated far field intensity at 783 nm for a $V_B^-$ emitter with in-plane emission dipole in unstructured hBN and integrated in a bullseye cavity, respectively. The intensity distribution in the unstructured hBN was multiplied by a factor of 100 to make it visible. f) Average over several far field measurements of $V_B^-$ from unstructured hBN. g) Far field measurement from the center of device $BE_3$.*

Figures 3(b) and 3(c) display the simulated electric field |E| for the in-plane dipole at the center of the bullseye cavity, showing a highly localized field at the center of the device, with a clear directionality and collimated emission to the perpendicular directions of the bullseye. This becomes more evidence in simulations of the far field intensity, compared for the emission from unstructured hBN and emission from a $V_B^-$ center coupled to a bullseye cavity in Figure 3(d) and (e), respectively.

Noteworthy, the far field emission from the color center inside the structure is highly directional with the intensity into low angles (10°) increased by a factor of ~ 2000. Figure 3 (f, g) show back focal plane images of the pristine hBN and the bullseye cavity, with an excellent matching field profile to the simulation. Due to this directionality effect, higher collection efficiencies can be expected with objectives of lower NA, yielding even higher experimental countrates (~ factors 10 - 100). Moreover, the directional emission from the bullseye cavity would be perfectly suitable for coupling into a single mode fiber, due to the close NA match, which is a requirement for integration towards several practical applications. We note that in support of our prior conclusion of the in plane dipole orientation, we observed in simulation that the far field intensity distribution for an out of plane dipole would not match the experimental results observed here (Supporting Information 3).

Finally, to showcase the enhanced spin readout properties in the bullseye cavity, we measure the ODMR signature of the $V_B^-$ ensembles at room temperature. The level scheme of the $V_B^-$ is shown in figure 4(a) and consists of a fully non-degenerate triplet ground state ($3A_2$'), with a zero field splitting of ~ 3.5 GHz between the $m_s$=0 and $m_s$=+/-1 states. For the ODMR measurements, a copper microwire (~ 20 μm diameter) was placed in the vicinity (few 10 μm) of the bullseye cavities, and the microwave was swept in a range from 3.1 to 3.7 GHz as the total PL counts were recorded with an avalanche photodiode. A PL map of the structure next to the wire is shown in Figure 4(b), where ODMR spectra were taken from the center of the bullseye cavity and outside of the structured region. Here, the PL counts from the center of the bullseye cavity are a factor of ~ 5 higher relative to unstructured hBN as directly apparent from the PL map. Here the high directionality of the emission becomes beneficial, because the microwire obstructs some of the collection into larger angles for emission from unstructured hBN. However, this is negligible for emitters inside the bullseye cavity, as the emission is highly directional (discussed in Figure 3).

A direct comparison of an average over 10 microwave sweeps is shown in Figure 4(a). Due to an improved signal-to-background ratio, the ODMR spectra of the VB- inside a bullseye cavity (Figure 4(c), upper panel) displays a contrast of ~ 5.4 % as compared to the signal from the pristine hBN (~ 3.6 %). Due to the improved PL counts from the bullseye cavity we also achieved a better signal-to-noise ratio, indicated by a lower value for the average standard deviation of 1.2 % vs. 1.7 %. The improved ODMR contrast and an improved signal-to-noise ratio is an important outcome that can yield single sweep readout and enable further integration of hBN devices with other 2D materials for sensing applications. We emphasize that the demonstration of an improved ODMR readout is not necessarily guaranteed. Specifically, as fabrication may introduce rough sidewalls and lattice distortion, which can affect the $V_B^-$ spins in a detrimental way by introducing non radiative decay paths that reduce the ODMR contrast. However, the results show clearly that the device structure is suitable and the fabrication did not affect the spin defect.

In summary we fabricated monolithic hBN bullseye cavities and demonstrated the first coupling of $V_B^-$ spin defects to a photonic cavity. We achieved a significant enhancement of the collected PL signal (6-fold and 10-fold for collection at ~ 60° and ~ 10° respectively). Using FDTD modelling, we further presented strong evidence that the emission dipole of the $V_B^-$ spin defects is in plane, as spectra and far-field images match almost perfectly. Ultimately, we proved the advanced functionality of the fabricated devices by showcasing an improved contrast and signal-to-noise ratio for ODMR measurements. Our results constitute an important step forward in employing hBN for integrated quantum photonic devices.

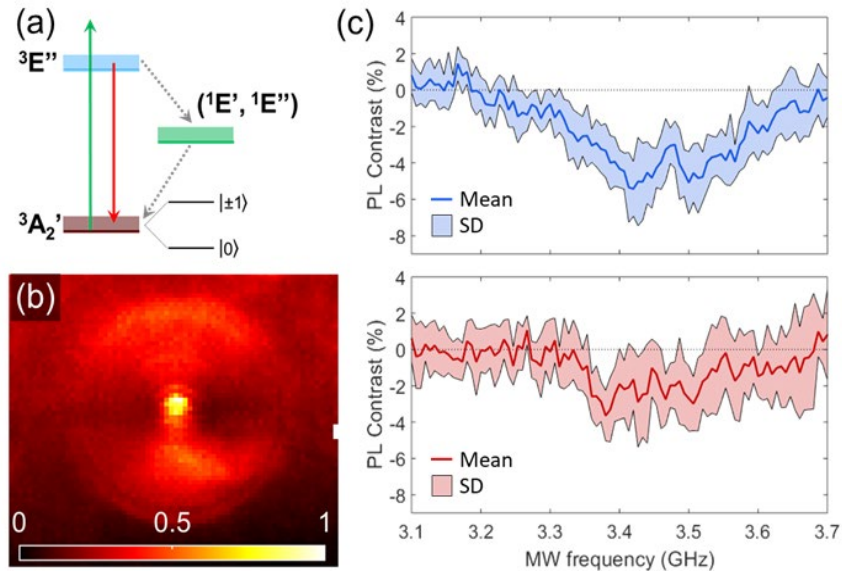

*Figure 4. Optically Detected Magnetic Resonance.* a) The level structure of the $V_B^-$ defect with a triplet ground state, indicated by 0 and +/- 1 states. b) PL map of the BE cavity taken for ODMR measurements. The color scale represents normalized PL counts. c) ODMR spectra of VB- inside (top panel) and outside of the cavity (bottom panel). The solid lines correspond to the mean over 10 scans, the shaded regions represent the standard deviation of the mean.

**Acknowledgement**

We acknowledge the Australian Research Council (CE200100010, DP190101058) and the Asian Office of Aerospace Research and Development (FA2386-20-1-4014). V. D. acknowledge financial support from the DFG through the Würzburg-Dresden Cluster of Excellence on Complexity and Topology in Quantum Matter—ct.qmat (EXC 2147, project-id 39085490)